# Superconductivity in repulsive Fermi-systems at low density.

M.Yu. Kagan[1], D.V. Efremov[2], M.S. Mar'enko[3], V.V. Val'kov[4].

[1]*Kapitza Institute for Physical Problems, Kosygin str. 2, 19334 Moscow, Russia*

[2]*Max -Planck-Institut für Festkörperforschung, D-70569 Stuttgart, Germany*

[3]*Department of Physics and Astronomy, Hofstra University, Hempstead, 11549 New York, USA*

[4]*Kirenskii Institute of Physics, 660036 Krasnoyarsk, Russia*

In the large variety of models such as 3D and 2D Fermi-gas model with hard-core repulsion, 3D and 2D Hubbard model, and Shubin-Vonsovsky model we demonstrate the possibility of triplrt p-wave pairing at low electron density. We show that the critical temperature of the p-wave pairing can be strongly increased in a spin-polarized case or in a two-band situation already at low density and reach experimentally observable values of (1-5)K. We also discuss briefly d-wave pairing and high-Tc superconductivity with Tc~100K which arises in the t-J model in the range of parameters realistic for cuprates.

## Introduction.

One of the most important questions in connection with the theory of HTSC is whether it is possible to convert the sign of Coulomb interaction between electrons [1]. The first attempt to answer this question in a positive way was made by Kohn and Luttinger in 1965 [2]. Unfortunately their $T_C$ was unrealistically small. Our answer is much more optimistic. We proved this statement at low density limit, where we are far from AFM and structural instabilities. Moreover in this limit we can develop regular perturbation theory. The small parameter in the problem is a gas parameter $ap_F$ ($a$ is the scattering length, $p_F$ is Fermi momentum).

The $T_C$ - values which we obtain are not very low. Moreover our theory often works even for rather high densities due to the intrinsic nature of superconductive instabilities. In the last case the superconductive temperatures are reasonable.

## The Fermi-gas model.

The basic model for our theory is a Fermi-gas model.

In the case of repulsive interaction between two particles in vacuum the scattering length $a > 0$. However, effective interaction in substance, which is formed via polarization of a fermionic background, contains attractive p-wave harmonic and hence the system is unstable towards triplet p-wave superconductive pairing below the temperature [3,4]:

$$T_{C1} \sim \varepsilon_F \exp\left\{-\frac{1}{(ap_F)^2}\right\}. \qquad (1)$$

Effective interaction in substance in first two orders of perturbation theory is given by:

$$V_{eff}(p,k) = ap_F + (ap_F)^2 \Pi(p+k), \quad (2)$$

where $\Pi(p+k)$ is an exchange diagram which coincides in the case of a short range interaction with polarization operator.

Besides a regular part it contains a Kohn's anomaly of the form (in the 3D case):

$$\Pi_{\sin g} \sim (\tilde{q} - 2p_F) \ln |\tilde{q} - 2p_F|, \qquad (3)$$

where $\tilde{q} = |\bar{p} + \bar{k}|$ is a transferred momentum in a crossed channel. As a result we start from pure hard-core repulsion in vacuum and obtain the competition between repulsion and attraction in substance. The singular part of $V_{eff}$ 'plays' in favor of attraction and the regular part in favor of repulsion. S-wave superconductivity is suppressed by hard core. However for $l \neq 0$ hard core is ineffective. Moreover already at $l = 1$ the attractive contribution is dominant. The exact solution yields [3-4]:

$$T_{Cl} - \varepsilon_F \exp\left\{-\frac{5\pi^2}{4(2\ln 2 - 1)(ap_F)^2}\right\} = \varepsilon_F \exp\left\{-\frac{13}{\lambda^2}\right\}, \qquad (4)$$

where $\lambda = \frac{2ap_F}{\pi}$ is an effective 3D gas-parameter of Galitskii [9].

## Two-dimensional case.

In 2D effective interaction in first two orders of the gas-parameter has a form [7,8]:

$$V_{eff}(q) \sim f_0 + f_0^2 \Pi(\tilde{q}); \; f_0 = \frac{1}{2\ln(p_F r_0)} \text{ - is 2D gas-parameter of Bloom [10].}$$
$$(5)$$

where $r_0$ is the range of the potential.

However $V_{\sin g}(q) \sim f_0^2 \, \text{Re} \sqrt{\tilde{q} - 2p_F} = 0$ for $q \leq 2p_F$ - the Kohn's anomaly has one-sided character and is ineffective for the superconductivity. SC appears only in the third order in $f_0$ [7,8] where we have $f_0^3 \, \text{Re} \sqrt{2p_F - \tilde{q}}$ for the singular contribution to $V_{eff}(q)$. Exact evaluation of all third order diagrams yield [7,8]:

$$T_{Cl} \sim \varepsilon_F \exp\left\{-\frac{1}{6.1 f_0^3}\right\}. \qquad (6)$$

# 3D and 2D Hubbard model. Shubin-Vonsovsky model.

The same results for p-wave critical temperature (4,6) are valid for 3D and 2D Hubbard models [17] with repulsion. For the Hubbard model 3D gas-parameter of Galitskii [9] reads $\lambda = \frac{2dp_F}{\pi}$ (where d is intersite distance) and 2D gas-parameter of Bloom [10] $f_0 = \frac{1}{2\ln\frac{1}{2p_F d}}$. In 2D Hubbard model at low electron density and weak-coupling case also $d_{xy}$ - pairing is realized [18]. We proved an existence of superconductivity in more than ten 2D and 3D models. In most of the models we obtained p-wave pairing including the most repulsive and the most unbeneficial for SC Shubin-Vonsovsky model [11]. The Hamiltonian of the Shubin-Vonsovsky model reads:

$$H = -t \sum_{\langle ij \rangle \sigma} c_{i\sigma}^+ c_{j\sigma} + U \sum_i n_{i\uparrow} n_{i\downarrow} + \frac{V}{2} \sum_{\langle ij \rangle} n_i n_j, \quad (7)$$

where U is onsite Hubbard repulsion and V is additional Coulomb repulsion on neighboring sites, t is hopping integral. An effective vacuum interaction for Shubin-Vonsovsky model has a form (see Fig. 1).

Even in the most repulsive strong-coupling limit of the model $U \gg V \gg W$ (W is the bandwidth; W=12t for 3D simple cubic lattice; W=8t for square lattice in 2D) we get the same critical temperatures of the p-wave pairing (4,6) as in the absence of additional Coulomb repulsion (for V=0) both in 3D and 2D-cases.

The additional Coulomb repulsion V changes only preexponential factors in (6) and (8) (see [12,13]). It is an important result in connection with the discussion about a possible role of long-range screened Coulomb interaction for non-phonon mechanisms of SC started in [33-35].

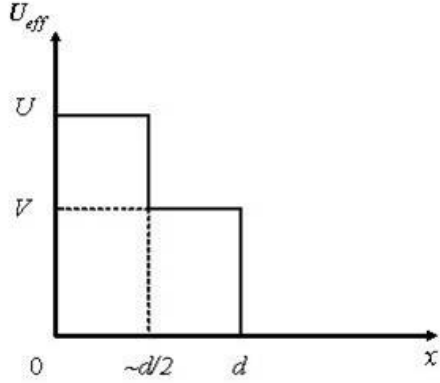

Fig. 1 Effective vacuum interaction in the Shubin-Vonsovsky model with Hubbard onsite repulsion $U$ and additional Coulomb repulsion $V$ on neighboring sites.

For higher densities of electrons there are Verwey localization [14-15] with checkerboard charge-ordered state in the strong-coupling limit of the model for dimensionless electron density $n_{el} = 1/2$ and Mott-Hubbard localization with an appearance of AFM-state [16,17] for $n_{el} = 1$. We also have here extended regions of phase separation close to $n_{el} = 1/2$ and $n_{el} = 1$ (see Fig. 2 and [19-21]). Thus our considerations for homogeneous SC in strong-coupling case $U >> V >> W$ are valid till the densities $n_{el} = 1/2 - \delta_C$, where for $V >> t$:

$$\delta_C \sim \left(\frac{t}{V}\right)^{1/2} \text{ in 2D and } \delta_C \sim \left(\frac{t}{V}\right)^{3/5} \text{ in 3D.} \quad (8)$$

For $\frac{1}{2} - \delta_C < n_{el} < \frac{1}{2}$ we have nano-scale phase-separation on small metallic clusters in the insulating checkerboard CO-matrix (see Fig.3).

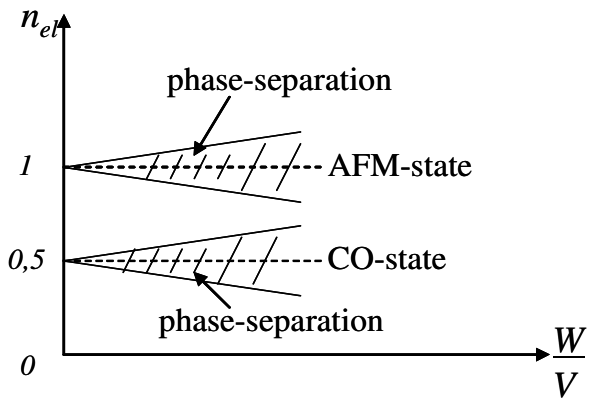

Fig.2 Qualitative phase-diagram of the Shubin-Vonsovsky model in the strong coupling case. At $n_{el} = 1$ AFM-state appears in the model, while at $n_{el} = 1/2$ we have the checkerboard CO-state. We have also extended regions of phase-separation close to $n_{el} = 1/2$ and $n_{el} = 1$

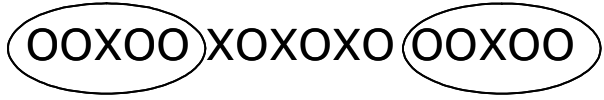

Fig.3 Phase-separated state at the densities $\frac{1}{2} - \delta_C < n_{el} < \frac{1}{2}$ with nano-scale metallic clusters inside CO checkerboard insulating matrix for $V \gg t$.

At critical concentrations $n_{el} = \frac{1}{2} - \delta_C$ the metallic clusters start to touch each other. As a result an infinite metallic cluster appears (all the sample volume becomes metallic) for $n_{el} < \frac{1}{2} - \delta_C$.

In the opposite Born case $W > U > V$ the phase-separation is absent in the model and we can construct SC phase-diagram for p-wave, $d_{xy}$ and $d_{x^2-y^2}$ - wave pairing for all the densities $0 < n_{el} < 1$. The first results in this case were obtained in [35].

## The possibility to increase $T_C$ already at low density

There are two possibilities to increase $T_C$ already at low density [5,23]:
to apply an external magnetic field (or to create strong spin-polarization)[5]
to consider a two-band situation[23].
In both cases the most important idea is an idea of separation of the channels. In magnetic field the Cooper pair is formed by two spins 'up' while effective interaction is prepared by two spins 'down'. As a result the Kohn's anomaly increases. For $H \neq 0$ it becomes:

$$\Pi_{\sin g}(q) \sim (q_\uparrow - 2p_{F\downarrow}) \ln | q_\uparrow - 2p_{F\downarrow} | = (\theta - \theta_C) \ln(\theta - \theta_C) \qquad (9)$$

and $\theta_C$ differs from $\pi$ proportionally to $\left( \frac{p_{F\uparrow}}{p_{F\downarrow}} - 1 \right)$. Thus already first derivative of $\Pi_{\sin g}$ and the effective interaction with respect to $(\theta - \theta_C)$ is divergent. Note that for $H = 0$ the Kohn's anomaly reads: $(\pi - \theta)^2 \ln(\pi - \theta)$ and only second derivative of $V_{eff}$ with respect to $(\pi - \theta)$ is divergent.

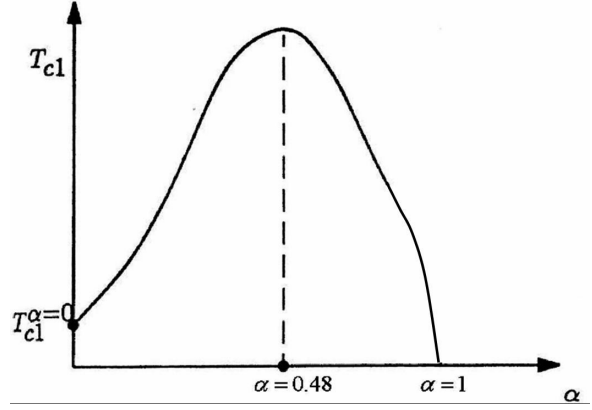

Fig.4 Polarization dependence of $T_C$ in 3D case.

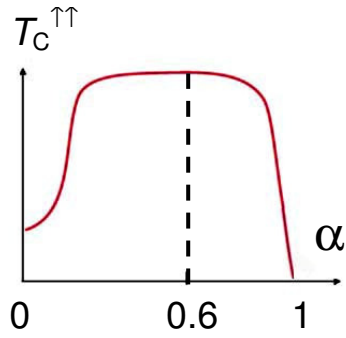

Fig. 5 Polarization dependence of $T_C$ in 2D case.

Unfortunately there is a competing process: namely the decrease of the density of states of the 'down' spins: $N_\downarrow(0) = \dfrac{m p_{F\downarrow}}{4\pi^2}$. As a result of this competition $T_C^{\uparrow\uparrow}$ has reentrant behavior with large maximum (see Fig. 4). This theory is confirmed by experiments of Frossati group in Leiden[51]: for $^3$He $T_C^{\uparrow\uparrow}(\alpha = 6\%) = 3.2mK$ while $T_C(\alpha = 0) = 2.7mK$. As a result we obtain 20% increase of critical temperature. In maximum $T_C^{\uparrow\uparrow} = 6.4 T_C$ for $^3$He and $T_C^{\uparrow\uparrow} = 10^5 T_C$ for mixtures [52].

In 2D films of $^3$He in a magnetic field we have $\Pi(q) \sim \mathrm{Re}\sqrt{q_\uparrow - 2 p_{F\downarrow}}$ and large 2D Kohn's anomaly becomes effective for superconductivity. The maximum is broad and very large (see Fig. 5) it stretches from $\alpha = 0,1$ till $\alpha = 0,9$. In maximum (for $\alpha = 0,6$):

$$T_{C\max}^{\uparrow\uparrow} = \varepsilon_F \exp\left\{ -2\ln^2\left( \frac{1}{p_F r_0} \right) \right\}. \qquad (10)$$

$T_C$ in maximum is bigger in 16 times in exponent then $T_C$ in 3D $T_{C\max}^{\uparrow\uparrow} \to \varepsilon_F e^{-2}$ for $\ln\left( \dfrac{1}{p_F r_0} \right) \to 1$.

The same result could be obtained in 2D electron gas in a parallel magnetic field [24]. Magnetic field does not change the motion of electrons in plane here. The Meissner effect is suppressed. Hence we have qualitatively the same situation as in uncharged (neutral) $^3$He films (see Fig. 6) and reentrant superconductive behavior for $T_C$ in field. For $H \sim 15$ Tesla and $\varepsilon_F \sim 30K$ $T_{C1} \sim 0.5K$ .

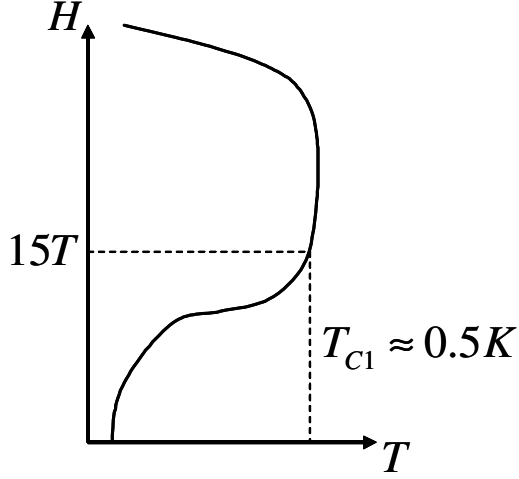

Fig. 6 H-T diagram for 2D electron gas in parallel magnetic field.

## The two-band Hubbard model

In two bands the role of spins 'up' play electrons of the first band while the role of spins 'down' – electrons of the second band. The connection between the bands is due to interband Coulomb interaction $U_{12}n_1n_2$ . The following excitonic mechanism of superconductivity is possible: the Cooper pairs are formed in one band due to polarizations of the second one [23,25,26].

The role of spin polarization $\alpha$ plays the relative filling of the bands $n_1 / n_2$ (see Fig. 7). If we consider the two-band Hubbard model with one narrow band, than an effective interaction is mostly governed by heavy-light repulsion (see Fig.8) and

$$T_{C\max} = T_C\left(\frac{n_h}{n_L} \approx 4\right) = \varepsilon_F \exp\left\{-\frac{1}{2f_0^2}\right\} \qquad (11)$$

where $n_1 = n_h$ for heavy band; $n_2 = n_L$ for light band; $U_{12} = U_{hL}$ - "heavy-light" interband Hubbard repulsion.

In Born weak-coupling case $f_0^2 = \dfrac{m_h m_L}{4\pi^2} U_{hL}^2$ depends upon interband Hubbard interaction $U_{hL}$ [23]. In strong-coupling case $f_0^2 = \dfrac{m_h}{m_L} \dfrac{1}{\ln^2 \dfrac{1}{p_F^2 d^2}}$ [25-26]. Finally in the so-called unitarian limit of screened Coulomb interaction $f_0 \rightarrow \frac{1}{2}$ and

$T_{C\max} \sim \varepsilon^*_{Fh} \exp\{-2\}$ [25-26], where renormalized Fermi-energy

$\varepsilon^*_{Fh} = \frac{p^2_{Fh}}{2m^*_h} \sim (30-50)K$ and enhanced heavy mass $m^*_h \sim 100m_e$ due to many-body

Electron-polaron effect [27-28]. As a result we can get $T_{C1} \sim 5K$ for Fermi-

energies $\varepsilon^*_{Fh} \sim (30-50)K$ - typical for uranium-based HF compounds. Note that

electron-polaron effect which produces strong heavy mass enhancement in this

model is connected with non-adiabatic part of the wave-function which describes

heavy electron dressed in the cloud of virtual electron-hole pairs of the light band

(see Fig. 9).

If we collect the polaron exponent we get [27-28]:

$$\frac{m^*_h}{m_h} = \left(\frac{m_h}{m_L}\right)^{\frac{b}{1-b}}, \qquad (12)$$

where $b = 2f_0^2$ in 2D and $b = 2\lambda^2$ in 3D. Hence for $f_0 = 1/2$ (unitarian limit of screened Coulomb interaction) $b = 1/2$; $\frac{b}{1-b} = 1$ and $\frac{m^*_h}{m_h} = \frac{m_h}{m_L}$. Correspondingly

$\frac{m^*_h}{m_L} = \left(\frac{m_h}{m_L}\right)^2$ and if we start with $\frac{m_h}{m_L} \sim 10$ in local density approximation (LDA-

scheme) [53], we can finish with $m^*_h \sim 100m_e$ due to many-body Electron-polaron effect and $T_{C1} \sim 5K$.

Thus we get an effective mass of heavy particles and superconductive temperatures realistic for uranium-based heavy fermion compounds.

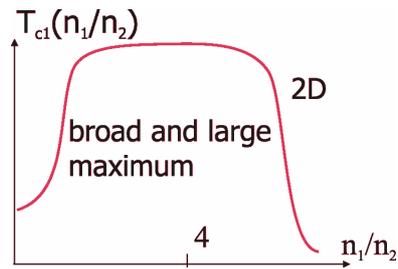

Fig. 7 $T_C$ as a function of relative filling in the two band model.

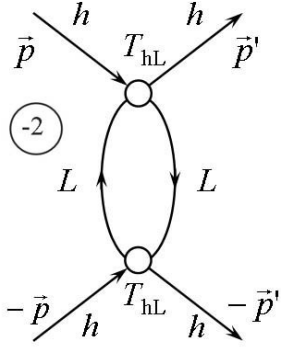

Fig.8 The leading contribution to the effective interaction $V_{eff}$ for the p-wave pairing of heavy particles via polarization of light particles. The open circles stand for the vacuum T-matrix $T_{hL}$, which in Born case coincides with interband Hubbard interaction $U_{hL}$.

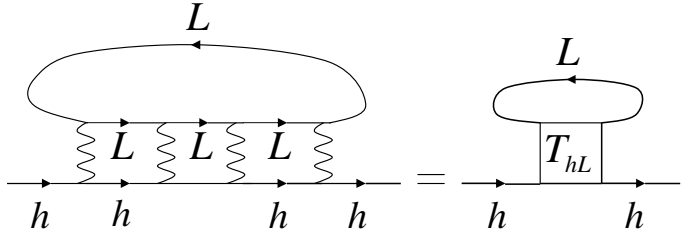

Fig.9 The lowest order skeleton diagram for EPE in the self-consistent T-matrix approximation. $T_{hL}$ stands for the T-matrix in substance on Fig.9

This mechanism can be important in Bi and T*l*-based HTSC-materials. It can also provide superconductivity in superlattices (PbTe-SnTe) and dichalcogenides (CuS$_2$, CuSe$_2$) with geometrically separated layers. Note that two bands also can belong to one layer. We suggested also that this mechanism could be dominant in Sr$_2$RuO$_4$ [12,25,26] and in fermionic $^6$Li in magnetic traps [22].

Note that in the case of one heavy and one light band with $m_h >> m_L$ and $n_h > n_L$ the critical temperature $T_C$ is mostly governed by pairing of heavy electrons via polarization of light electrons (see Fig. 8). However an inclusion of already infinitely small Geilikmann-Moskalenko-Suhl term $K\sum_{pp'}a_p^+a_{-p}^+b_{p'}b_{-p'}$ [38-42] which rescatters the Cooper pair between the two bands provides the opening of SC gaps in both heavy and light band at the same temperature.

## 2D t-J model.

We consider the 2D t-J model with released constraint [29,30]. The Hamiltonian reads:

$$H = -t\sum_{<ij>\sigma} c_{i\sigma}^+ c_{j\sigma} + U\sum_i n_{i\uparrow}n_{i\downarrow} + J\sum_{<ij>}(\vec{S}_i\vec{S}_j - \frac{1}{4}n_i n_j). \qquad (13)$$

It is a model with strong repulsion on one site and small AFM attraction J>0 on neighboring sites, i.e. $U >> \{J,t\}$ (Fig.10). The phase diagram of this model is given in Fig.11.

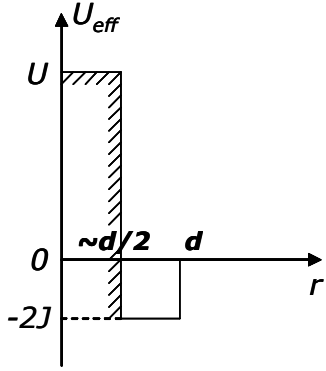

Fig. 10. Effective interaction in the t-J model with released constraint. U is onsite Hubbard repulsion, J is weak AFM-attraction on neighboring sites.

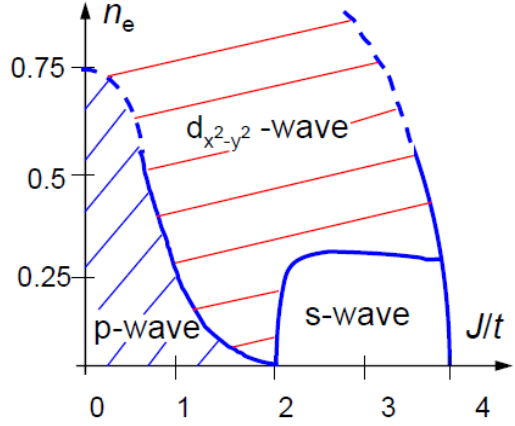

Fig. 11. Phase diagram of the 2D t-J model at small and intermediate densities

For the parameters realistic for optimally doped HTSC-materials J/t~1/2, $n_e = \dfrac{2\varepsilon_F}{W} = 0.85$ we have the following estimate for the critical temperature:

$$T_C^{x^2-y^2} \sim \varepsilon_F \exp\{-\frac{\pi t}{2Jn_e^2}\} \sim 10^2 K . \quad (14)$$

Note that the same estimate for $T_C$ of d-wave pairing ($d_{x^2-y^2}$) was obtained in a more rigorous theory of Plakida et al [31-32] for optimally doped cuprates. In underdoped case there is a possible bosonic motive and we could think about a BCS-BEC crossover [43-44] for pairing of two spin-polarons [49-50] (two AFM strings [45,46] or two composite holes each one containing spinon and holon[47,48] in the $d_{x^2-y^2}$ channel).

# Conclusion

On a large variety of models we proved an existence of p-wave pairing in purely repulsive fermion systems. We demonstrated the possibility to increase $T_C$ till experimentally feasible values ~5K already at low density in strongly spin-polarized case or in the two-band situation. The systems where triplet p-wave pairing is realized or can be expected include superfluid $^3$He, ultracold Fermi-gasses in the regime of p-wave Feshbach resonance [54], heavy-fermion superconductors such as $U_{1-x}Th_xBe_{13}$ and ruthenates $Sr_2RuO_4$, organic superconductor $\alpha$-(BEDT-TTF)$_2$I$_3$ and layered dichalcogenides $CuS_2$-$CuSe_2$, semimetals and semimetallic superlattices InAs-GaSb, PbTe-SnTe. Regarding possible high-$T_C$ superconductivity we demonstrated a simple estimate to get $T_C$ in the range of 100K of the d-wave pairing ($d_{x^2-y^2}$) in the parameter region typical for optimally doped cuprates in the framework of the 2D t-J model with released constraint.

We analyzed also the normal state of the basic models with repulsion and find the nontrivial corrections to Galitskii-Bloom Fermi-gas expansion due to the presence of the antibound state [36] in the lattice models or the singularity in Landau quasiparticle f-function at low density in 2D[37]. These corrections however, do not destroy Landau Fermi-liquid picture both in 3D and 2D.

Acknowledgements.

The authors acknowledge helpful discussions with A.V. Chubukov, A.S. Alexandrov, V.V. Kabanov, K.I. Kugel, Yu.V. Kopaev, N.M. Plakida, N.V. Prokof'ev. M.Yu.K. work was supported by RFBR grant №

# References.